\documentstyle[preprint,aps,pre,psfig]{revtex}
\newcommand{\be}{\begin{equation}}
\newcommand{\ee}{\end{equation}}
\newcommand{\ba}{\begin{eqnarray}}
\newcommand{\ea}{\end{eqnarray}}
\newcommand{\bsigma}{\mbox{\boldmath $\sigma$}}

\newcommand{\eps}{\epsilon}
\newcommand{\bk}{{\bf k}}
\newcommand{\br}{{\bf r}}
\input amssym.def
\newsymbol\lesssim 132E
\begin{document}
\title{Ring kinetic theory for
an idealized granular gas}
\author{T.P.C. van Noije and M.H. Ernst\\
{\it Instituut voor Theoretische Fysica, Universiteit Utrecht,
Postbus 80006,
3508 TA Utrecht, The Netherlands}}
\date{\today}
\maketitle
\begin{abstract}
The dynamics of {\em inelastic} hard spheres is described in terms of 
the binary collision expansion, yielding the corresponding 
pseudo-Liouville equation and
 BBGKY hierarchy for the reduced distribution functions. 
Based on cluster expansion techniques we derive the Boltzmann and
ring kinetic equations for inelastic hard spheres.
In the {\em simple ring} approximation, we calculate the 
structure factor of vorticity fluctuations in a freely evolving, 
dilute granular gas. 
The kinetic theory result agrees with the result, derived previously 
from fluctuating hydrodynamics.
In the limit of incompressible flow, this structure factor alone 
determines the spatial velocity correlations, which are of dynamic
origin and include long range $r^{-d}$-behavior.
The analytic results are compared with MD simulations.
\end{abstract}
\section{Introduction}
In recent years the interest in static and rheological properties of 
assemblies of mesoscopic or macroscopic bodies or granules has been 
rapidly increasing.
Both as a solid and as a fluid, granular systems show unusual
behavior \cite{jaeger}.
Flows of granular material can be subdivided into quasi static (contact)
flows and {\em rapid} (collision driven) {\em granular flows}.
In this characterization of Ref.\ \cite{jaeger}, the former is referred
to as the granular liquid regime and the latter as the {\em granular
gas regime}.
Only in rapid granular flows, the dynamics can be described by
sequences of binary collisions and, as a consequence, the
methods of kinetic theory are most suitable
\cite{lun,jenkins+richman,brey+moreno+dufty,brey+dufty+santos,goldshtein,goldhirsch,mcnamara,deltour,esipov,boston}. 
Such flows do obey
the standard conservation laws of mass and momentum, and can
therefore be considered as fluids.
However, energy is not conserved.
The dynamics is essentially dissipative, which gives rise to
several interesting new phenomena, such as clustering and inelastic
collapse \cite{jaeger}.

In rapid granular flows, collisions of granules are accompanied by 
conversion 
of kinetic energy into rotational energy and into energy of other 
internal degrees 
of freedom, and lead to effective `cooling'  phenomena in unforced flows.
Such flows have been extensively modeled through smooth and rough hard 
spheres with 
inelastic collisions \cite{lun,jenkins+richman,brey+moreno+dufty,brey+dufty+santos,goldshtein,goldhirsch,mcnamara,deltour,esipov,boston}.

Up to now the theoretical description of rapid granular flows has 
been based on Boltzmann-Enskog kinetic equations for hard sphere-type 
interactions.
Inherent to such descriptions is the molecular chaos assumption of 
uncorrelated binary collisions.
To study the extent to which collective effects may be of importance 
at a fundamental level of description, one needs to correct for the 
breakdown of the molecular chaos assumption and to account for the 
effects of dynamic correlations.

In the last 35 years many-body theories have been developed to account 
for these dynamic correlations in systems of microscopic particles 
obeying the standard conservation laws. 
The fundamental concept to describe these dynamic correlations are 
`ring collisions', i.e.\
sequences of correlated binary collisions, which lead to the so 
called {\rm ring kinetic theory}.
This ring kinetic theory for systems of perfectly smooth elastic hard 
spheres has been at the basis of all major developments in
nonequilibrium statistical mechanics over the last three
decennia:
it explains the logarithmic density dependence of the transport
coefficients, and the breakdown of the
virial expansion for transport coefficients 
\cite{brush,vleeuwen+weyland,vbeijeren+dorfman(berne)};
it explains the algebraic long time tails of the velocity autocorrelation
function and similar current-current correlation functions
\cite{alder,tails,dorfman+cohen}, 
the non-analytic dispersion relations for sound propagation
and for relaxation of hydrodynamic excitations
\cite{ernst-physicaD}, 
as well as the breakdown of the Navier-Stokes equations in
two-dimensional fluids at very long times and the non-existence of
linear transport coefficients in 2-$D$
\cite{resibois+pomeau};
moreover, it explains the existence of long range spatial
correlations in nonequilibrium {\em stationary states} 
\cite{d+k+s,schmittmann+zia}, driven by reservoirs which impose shear
rates or temperature gradients, or in driven diffusive systems.
Such systems violate the conditions of detailed balance and the
stationary states are non-Gibbsian states \cite{bussemaker}.

The goal of this paper is to explain the {\em long range}
correlations (in fact, {\em intermediate} range, as the algebraic tails
are exponentially cut off at the largest scales), observed in 
molecular dynamics simulations of unforced flows of two-dimensional 
granular gases
\cite{boston,noije+ernst+brito+orza,brito+orza+noije+ernst}.
This will be done by extending
the mean field-type Boltzmann-Enskog 
equation for rapid granular flows by including ring collisions.
To do so we choose the model of perfectly smooth, but inelastic hard 
disks or spheres ($d=2,3$) of diameter $\sigma$. 
The model incorporates the most fundamental feature of the dissipative 
dynamics of granular flow, namely the conversion of kinetic energy into 
internal energy.
Moreover, it is a many-body system with well-defined and relatively 
simple dynamics, which has already widely been used in molecular dynamics
simulations \cite{goldhirsch,mcnamara,deltour,esipov,boston,noije+ernst+brito+orza,brito+orza+noije+ernst}. 

The interactions between smooth {\em inelastic hard spheres} (IHS) only 
affect the translational degrees of freedom and are modeled by
instantaneous collisions as in the case of elastic hard spheres.
During a collision momentum will be transferred along the line joining 
the centers of mass of the two colliding particles, indicated by the 
vector ${\bsigma}$ pointing from the center of particle 2 to that of 
particle 1, as illustrated in Fig.\ 1.

Upon collision, inelastic hard spheres of equal mass $m$ 
change their velocities according to the collision rules
\ba
{\bf v}^{\ast}_1&=&{\bf
v}_1-\textstyle{\frac{1}{2}}(1+\alpha)({\bf
v}_{12}\cdot\hat{\bsigma})\hat{\bsigma}\nonumber\\
{\bf v}^{\ast}_2&=&{\bf
v}_2+\textstyle{\frac{1}{2}}(1+\alpha)({\bf
v}_{12}\cdot\hat{\bsigma})\hat{\bsigma},
\label{eq:collr} 
\ea
where $\hat{\bsigma}=\bsigma/\sigma$ denotes a unit vector,
${\bf v}_{12}={\bf v}_1-{\bf v}_2$, and 
$\alpha$ is the 
coefficient of normal restitution. 
Whereas the total momentum of the two particles is conserved in a
collision, the
total energy loss is $\textstyle{\frac{1}{4}}m \eps ({\bf
v}_{12}\cdot\hat{\bsigma})^2$, where the coefficient of inelasticity
is $\eps=1-\alpha^2$.

The restituting (precollision) velocities $({\bf v}_1^{\ast\ast},{\bf
v}_2^{\ast\ast})$ giving rise 
to $({\bf v}_1,{\bf v}_2)$, are found by inverting collision rule 
(\ref{eq:collr}) 
and given by (see Fig.\ 1)
\ba
{\bf v}^{\ast\ast}_1&=&{\bf v}_1-\textstyle{\frac{1}{2}}(1+\alpha^{-1})
({\bf v}_{12}\cdot\hat{\bsigma})\hat{\bsigma}\nonumber\\
{\bf v}^{\ast\ast}_2&=&{\bf v}_2+\textstyle{\frac{1}{2}}(1+\alpha^{-1})
({\bf v}_{12}\cdot\hat{\bsigma})\hat{\bsigma}.
\label{eq:revcollr}
\ea
Note that this inversion is not possible if $\alpha= 0$, which is a 
trivial limit.

The kinetic theory for inelastic hard spheres will be developed in close 
analogy with that for elastic ones.
The rather singular streaming operators $S_t(\Gamma)$, which
generate the $\Gamma$-space trajectories of the $N$-hard sphere
system, are reformulated (see section \ref{sec:BCE})  
in terms of $T$-operators (binary collision operators), following
the original derivation in Ref.\
\cite{ernst+dorfman+hoegy+vleeuwen}.
This allows us to introduce
a pseudo-Liouville equation for the
$N$-particle $\Gamma$-space distribution function, and obtain the
Bogoliubov, Born, Green, 
Kirkwood and Yvon (BBGKY) hierarchy for the reduced distribution 
functions.
The Liouville equation or the BBGKY hierarchy forms the standard starting 
point for deriving kinetic equations, such as the Boltzmann equation 
and the ring kinetic equation. This program is carried out in section 
\ref{sec:BBGKY}.
Subsequently, we use the ring kinetic equation for inelastic hard spheres
to calculate the nonequilibrium pair distribution function in 
a freely evolving, dilute granular gas.

A particular solution of the Boltzmann equation for this system is
the so called {\em homogeneous cooling state} (HCS), where the velocity
field vanishes everywhere, and the density $n$ and temperature $T(t)$
are spatially homogeneous,
while the temperature decays in time.
Some basic features of this HCS-solution are summarized in section
\ref{sec:HCS}. 
They are necessary
for an understanding of the remaining part of this paper, including its 
instability
against spatial fluctuations in flow velocity and density.
In section \ref{sec:corrtheory} we will calculate the build-up of
vorticity correlations using the ring kinetic equation.
The resulting expressions for the spatial correlations
$\langle u_\alpha(\br+\br^\prime,t) u_\beta(\br^\prime,t)\rangle$
in the flow field,
including long range $r^{-d}$-behavior, are 
identical to the results from fluctuating
hydrodynamics for incompressible velocity fluctuations
\cite{noije+ernst+brito+orza} in the low density limit.
The incompressibility assumption gives a valid description of the  
velocity correlations up to a length scale that diverges as
$1/\eps$, beyond which the $r^{-d}$-tail is cut off exponentially.
Section \ref{sec:corrtheory} concludes with a comparison of the 
theoretical predictions  
with the results of
molecular dynamics simulations of unforced flows in a
two-dimensional gas of inelastic hard disks (see
\cite{goldhirsch,mcnamara,brito+orza+noije+ernst}).

\section{The binary collision expansion}
\label{sec:BCE}
Having defined the dynamics of the system, we now turn to the statistical 
mechanics.
The ensemble average of a dynamical quantity $A(\Gamma)$
obeys the equality
\be
\int {\rm d}\Gamma \rho(\Gamma,0) A(\Gamma(t))=
\int {\rm d}\Gamma A(\Gamma) \rho(\Gamma,t),
\label{eq:av}
\ee
where
$\Gamma=\{x_1,x_2,\dots x_N\}$ with $x_i=\{{\bf r}_i,{\bf v}_i\}$
is a 
point in the $N$-particle phase space.
On the left hand side the time dependence is assigned to the 
dynamical 
variable $A(\Gamma(t))\equiv S_t(\Gamma)A(\Gamma)$ with 
$S_t(\Gamma)$ the time evolution operator, and on the right hand
side to the $N$-particle distribution function $\rho(\Gamma,t)$,
which can be done by
considering the terms in (\ref{eq:av}) as an inner product.
The time 
evolution of the $N$-particle distribution function is then given by
\be
\rho(\Gamma,t)=S^\dagger_t \rho(\Gamma,0), 
\label{eq:timedisp}
\ee
where $S^\dagger_t(\Gamma)$ is the adjoint of $S_t(\Gamma)$.
For Hamiltonian dynamics $S_t^\dagger(\Gamma)=S_{-t}(\Gamma)$.
 
For particles with hard core interactions the dynamics is undefined 
for physically inaccessible configurations, where the particles are 
overlapping. 
Such configurations have a vanishing weight in Eq.\ (\ref{eq:av}).
Since $S_t(\Gamma)$ only appears in the combination $\rho(\Gamma,0) 
S_t(\Gamma)$ which vanishes for overlapping initial configurations,
it suffices to consider $W_N(\Gamma)S_t(\Gamma)$, with 
$W_N(\Gamma)=\prod_{i<j}W(r_{ij})$ where $W(r)$ is the overlap
function: 
\be
W(r)=\left\{ \begin{array}{ll}
             0 & \mbox{if $r<\sigma$ (overlapping)} \\
             1 & \mbox{if $r>\sigma$ (non-overlapping).}
                    \end{array}
            \right.
\ee
However, the methods of many-body theory require formal perturbation 
expansions and subsequent resummations.
To do so, the time evolution operator $S_t(\Gamma)$ needs to be defined 
for {\em all} configurations, including the unphysical overlapping 
configurations.
A convenient representation, defined in all points in phase space, has 
been developed for elastic hard spheres in Ref.\ 
\cite{ernst+dorfman+hoegy+vleeuwen}, and is 
based on the binary collision expansion of $S_t(\Gamma)$ in terms of 
binary collision operators $T(ij)$.

The binary collision operator may be defined in terms of two-body 
dynamics through the time displacement operator $S_t(12)$, given in 
Ref.\ \cite{ernst+dorfman+hoegy+vleeuwen} by
\be
S_t(12)=S_t^0(12)+\int_0^t {\rm d}\tau S_{\tau}^0 (12) T(12) S_{t-\tau}^0 
(12).
\label{eq:st12}     
\ee
Here the free streaming operator $S_t^0(12)=\exp[t L^0(12)]$, 
with $L^0(12)=L^0_1+L^0_2$ and $L^0_i 
= {\bf v}_i\cdot \partial/\partial{\bf r}_i$. 
Its action on an arbitrary dynamical
function is
\be
S_t^0(12)A({\bf r}_1,{\bf p}_1,{\bf r}_2,{\bf p}_2)=A({\bf r}_1+
{\bf v}_1
t,{\bf p}_1, {\bf r}_2+{\bf v}_2 t, {\bf p}_2).
\ee
Following the argument of Ref.\ \cite{ernst+dorfman+hoegy+vleeuwen} for 
the case of elastic hard spheres step by step, the binary collision 
operator $T(12)$ for inelastic hard spheres is constructed as
\be
T(12)=\sigma^{d-1}\int_{{\bf v}_{12}\cdot\hat{\sigma}<0} {\rm d}
\hat{\bsigma}
|{\bf v}_{12}\cdot\hat{\bsigma}|\delta({\bf r}_{12}-{\bsigma})
(b^{\ast}_{\sigma}-1),
\ee
where $b^{\ast}_\sigma$ is an operator that replaces all (precollision)
velocities 
${\bf v}_i\,\, (i=1,2)$ appearing to its right by postcollision 
velocities 
${\bf v}_i^{\ast}$, defined for the inelastic case through collision
rule (\ref{eq:collr}), and $d$ is the dimensionality of the system.

The binary collision operator is defined for overlapping and 
non-overlapping configurations of two hard spheres. 
It extends the definition of $S_t(12)$ to all points in phase space.
In the ensemble average considered in Eq.\ (\ref{eq:av}), the overlap 
function $W_N(\Gamma)$ contains a factor $W(r_{12})$ which vanishes 
whenever $r_{12}<\sigma$.
Moreover, the pseudo-dynamics introduced through Eq.\ (\ref{eq:st12}) 
is noninvertible (see \cite{ernst+dorfman+hoegy+vleeuwen}).
Consequently, the generator $S_t(12)$ for two-particle dynamics is only 
defined for {\em positive} times.
For later reference we quote the property
$T(12)A(12)=0$,
where $A(12)$ is a constant or a function of the argument $({\bf v}_1
+{\bf v}_2)$, because of conservation of particle number and 
linear momentum in binary collisions. 
As kinetic energy is not conserved in inelastic collisions, this 
property does not apply if $A(12)$ is a function of 
the argument $(v_1^2+v_2^2)$.

Combinations of $T$-operators and free streaming operators $S_t^0$,
preceded by appropriate combinations of overlap functions, can be
used to construct the time displacement operators $S_t$ for dynamical
variables.
However to discuss the Liouville equation and describe the time
evolution of the reduced distribution functions we need to
consider the adjoint time displacement operators.

We start with the two-particle streaming operator $S_t(12)$.
In order to obtain the adjoint $S_t^\dagger(12)$, we consider the integral
equality
\be
\int {\rm d} x_1 {\rm d} x_2 B(12) W(12) S_t(12) A(12)=
\int {\rm d} x_1 {\rm d} x_2 A(12) 
S_{t}^{\dagger}(12) W(12) B(12),
\label{eq:inteq}
\ee
where ${\rm d} x_1 {\rm d} x_2={\rm d}{\bf r}_1 {\rm d}{\bf v}_1 
{\rm d}{\bf r}_2 {\rm d}{\bf v}_2$ and 
$A(12)$ and $B(12)$ are arbitrary functions of the phases $x_1$ and $x_2$.
Since $W(12)$ is appearing to the left of $S_t(12)$, this integral is 
well-defined.
Substituting $S_t(12)$ from Eq.\ (\ref{eq:st12}) and using Liouville's 
theorem $S_{-t}^{0\dagger}(12)=S_{t}^0(12)$ for free particle motion, 
the left hand side of Eq.\ (\ref{eq:inteq}) can be written as
\ba
\int {\rm d} x_1 {\rm d} x_2\Big\{&& A(12)S_{-t}^0(12)W(12)B(12)+ 
\nonumber\\
&&\int_0^t {\rm d}\tau [S_{-\tau}^0(12)W(12)B(12)] 
T(12)S_{t-\tau}^0(12)A(12)\Big\}.
\label{eq:intexpr}
\ea
Now notice that the binary collision operator $T(12)$ contains two terms:
a real collision term with $b^{\ast}_{\sigma}$ and a virtual collision 
term.
The real collision term appearing in (\ref{eq:intexpr}) can be 
transformed using 
${\bf v}_{12}^{\ast}\cdot \hat{\bsigma}=-\alpha {\bf v}_{12}\cdot
\hat{\bsigma}$, yielding a Jacobian ${\rm d}{\bf v}_1{\rm d}{\bf
v}_2={\rm d}{\bf v}_1^{\ast} {\rm d}{\bf v}_2^{\ast}/\alpha$,
and then relabeling ${\bf v}_i^{\ast}\rightarrow {\bf v}_i$ and 
${\bf v}_i\rightarrow {\bf v}_i^{\ast\ast}$.
In the virtual collision term we relabel $\hat{\bsigma}\rightarrow -
\hat{\bsigma}$ and use the free particle Liouville theorem to finally 
write (\ref{eq:intexpr}) in terms of the adjoint time displacement 
operator, defined through the right hand side of Eq.\
(\ref{eq:inteq}).
The resulting expression for the adjoint time displacement
operator becomes
\be
S_t^{\dagger}(12)=S_{-t}^0(12)+\int_0^t {\rm d}\tau S_{-\tau}^0(12)
\overline{T}(12)S_{-(t-\tau)}^0(12),
\ee
where the binary collision operator $\overline{T}$ is defined as
\be
\overline{T}(12)=\sigma^{d-1}\int_{{\bf v}_{12}\cdot\hat{\sigma}>0} 
{\rm d}\hat{\bsigma} ({\bf v}_{12}\cdot\hat{\bsigma})\left(
\frac{1}{\alpha^2}
\delta({\bf r}_{12}-\bsigma)b^{\ast\ast}_{\sigma}-\delta({\bf r}_{12}+
\bsigma)\right).
\label{eq:tmin}
\ee
Here $b^{\ast\ast}_{\sigma}$ acts on the velocities ${\bf v}_i\,\, 
(i=1,2)$ to its right and replaces them by restituting ones, 
${\bf v}_i^{\ast\ast}$, as defined in collision rule 
(\ref{eq:revcollr}).

A property of $\overline{T}$-operators, equivalent to $T(12)A(12)=0$,
is 
\be
\int {\rm d} x_1 {\rm d} x_2 A(12) \overline{T}(12) B(12) =0,
\label{eq:ab}
\ee
if $A(12)$ is a constant or a function of the argument 
$({\bf v}_1+{\bf v}_2)$,
whereas $B(12)$ is an arbitrary function.

The time displacement operators $S_t(12)$ can be put in a more convenient
form by using the property, $T(12)S_t^0(12)T(12)=0$,
valid for any $t>0$.
It also holds with $T$ replaced by $\overline{T}$.
This relation expresses the fact that two hard spheres cannot collide 
more than once with only free propagation in between.
Using this property, the time displacement operators can be written as
\ba
W(12)S_t(12)=W(12)\exp[tL^0(12)+ tT(12)]\nonumber\\
S_t^{\dagger}(12)W(12)=\exp[-tL^0(1,2)+ t\overline{T}(12)]W(12).
\ea
 
We now return to the full $N$-particle system.
As shown in Ref.\ \cite{ernst+dorfman+hoegy+vleeuwen}, 
the dynamics of the $N$-particle system 
can be represented in the compact form of pseudo-streaming operators 
with the help of the above property $T(12)S_t^0(12)T(12)=0$,
yielding
\ba
W_N(\Gamma)S_t(\Gamma)=W_N(\Gamma)\exp[tL^0(\Gamma)+ t\sum_{i<j}T(ij)]
\nonumber\\
S_t^{\dagger}(\Gamma)W_N(\Gamma)=\exp[-tL^0(\Gamma)+t\sum_{i<j}
\overline{T}(ij)]W_N(\Gamma),
\label{eq:sdagger}
\ea
with $L^0(\Gamma)=\sum_i L^0_i$ the free particle streaming
operator.
These time evolution operators are defined everywhere in phase space, 
and the overlap function gives a vanishing weight to unphysical 
configurations, provided that
$W_N(\Gamma)$ always appears to the left of $T$-operators, 
or to the right of $\overline{T}$-operators.

A minor generalization is to include a conservative external force field 
in the dynamics.
In that case the single-particle free streaming operator should be 
defined as
\be
L^0_i={\bf v}_i\cdot \frac{\partial}{\partial {\bf r}_i} + {\bf a}_i 
\cdot
\frac{\partial}{\partial {\bf v}_i},
\ee
where ${\bf a}_i$ is the external force per unit mass, acting on the 
$i$-th particle.

Similar results for the binary collision operators $T$ and
$\overline{T}$ for inelastic hard spheres have been derived
independently by Brey et al.\ 
\cite{brey+dufty+santos}.

\section{BBGKY hierarchy}
\label{sec:BBGKY}
The time evolution of the $N$-particle distribution function 
$\rho(\Gamma,t)=\rho(x_1,x_2,\dots x_N,t)$, with $x_i=\{{\bf r}_i,
{\bf v}_i\}$ is given by the Frobenius-Perron equation.
For conservative Hamiltonian systems this equation is the Liouville 
equation which is an expression of the incompressibility of the flow. 
In the case of dissipative systems which by definition are time 
irreversible, the phase space volumes are contracted along the flow.
According to Eqs.\ (\ref{eq:timedisp}) and (\ref{eq:sdagger})  the time 
evolution of the distribution function for inelastic hard spheres is 
given by the pseudo-Liouville equation
\be
[\partial_t+ L^0(\Gamma)]\rho(\Gamma,t)= 
\sum_{i<j} \overline{T}(ij) \rho(\Gamma,t).
\label{eq:pseudo}
\ee
An equivalent representation of the time evolution of the system can
be given 
in terms of reduced $s$-particle distribution functions ($s=1,2,\dots$), 
defined as
\be
f_{12\dots s}(t)\equiv f^{(s)}(x_1,x_2,\dots x_s,t)=\frac{N!}{(N-s)!}
\int {\rm d} x_s\dots {\rm d} x_N\rho(x_1,x_2,\dots x_N,t),
\label{eq:reduced}
\ee
where $\rho(t)$ is normalized to unity. 
Integration of Eq.\ (\ref{eq:pseudo}) using (\ref{eq:ab}) yields the 
BBGKY hierarchy for the reduced distribution functions.
We only quote the first two hierarchy equations:
\ba 
(\partial_t+ L^0_1)f_1&=&
\int {\rm d} x_2 \overline{T}(12) f_{12}\nonumber\\
\big[ \partial_t+ L^0_1+ L^0_2 -  \overline{T}(12)\big] f_{12}
&=& \int {\rm d} x_3 [ \overline{T}(13)+ \overline{T}(23)] f_{123}.
\label{eq:bbgky}
\ea
This set of equations is an open hierarchy, which expresses the time 
evolution of the $s$-particle distribution function in terms of the 
$(s+1)$-th function.

In the literature on kinetic theory of inelastic hard spheres the first 
equation of the BBGKY hierarchy has frequently been derived intuitively 
and used as a starting point to obtain the Boltzmann-Enskog equation for 
the single-particle distribution function 
\cite{lun,jenkins+richman,brey+dufty+santos,goldshtein}.
Using the explicit expression (\ref{eq:tmin}) for $\overline{T}(12)$ it 
can be written in full detail as
\ba
&&(\partial_t + L^0_1)
f({\bf r}_1,{\bf v}_1,t)=
{\sigma^{d-1}}\int {\rm d}{\bf v}_2 \int_{{\bf v}_{12}\cdot\hat{\sigma}>0} 
{\rm d}\hat{\bsigma} ({\bf v}_{12}\cdot\hat{\bsigma})\times\nonumber\\
&&\left\{\frac{1}{\alpha^2}f^{(2)}({\bf r}_1,{\bf v}_1^{\ast\ast},
{\bf r}_1-
{\bsigma},{\bf v}_2^{\ast\ast},t)-f^{(2)}({\bf r}_1,{\bf v}_1,{\bf r}_1+
{\bsigma},{\bf v}_2,t)\right\}.
\label{eq:enskog}
\ea

The corresponding equation for the rate of change of an average is 
obtained by multiplying the first hierarchy equation in (\ref{eq:bbgky}) 
with $\int {\rm d}{\bf v}_1 \psi({\bf v}_1)$, and using the adjoint of 
$\overline{T}(12)$ to find
\ba
&&\frac{\partial}{\partial t}\int {\rm d}{\bf v}_1 \psi({\bf v}_1) 
f({\bf r}_1,
{\bf v}_1,t) + \frac{\partial}{\partial {\bf r}_1} \cdot \int {\rm
d}{\bf v}_1 
{\bf v}_1 \psi({\bf v}_1) f({\bf r}_1,{\bf v}_1,t) \nonumber\\
=&& \int {\rm d}{\bf v}_1 \int {\rm d} x_2 f_{12} T(12) \psi({\bf v}_1)
\nonumber\\
=&& \sigma^{d-1} \int {\rm d}{\bf v}_1 \int {\rm d}{\bf v}_2 
\int_{{\bf v}_{12}\cdot
\hat{\sigma}>0} {\rm d}\hat{\bsigma} ({\bf v}_{12}\cdot\hat{\bsigma})\times 
\nonumber\\
&& f^{(2)}({\bf r}_1,{\bf v}_1,{\bf r}_1+\bsigma,{\bf v}_2,t) [\psi(
{\bf v}_1^\ast)-\psi({\bf v}_1)],
\label{eq:ratepsi}
\ea
where we have set the external force equal to zero.
This equation is the starting point for deriving macroscopic conservation 
laws, hydrodynamic equations and the rate of change of the temperature.

We return again to the kinetic equations.
In order to derive a closed equation for the single-particle distribution 
function $f$ or for the pair function $f_{12}$ some kind of closure 
relation is required, to express $f_{12}$ in terms of $f$, such as 
Boltzmann's molecular chaos assumption, 
Bogoliubov's functional assumption \cite{cohen-nijenrode} or  
cluster expansion methods 
\cite{vbeijeren+dorfman(berne),cluster}.
Here we shall illustrate how the methods derived in the kinetic theory 
for elastic hard spheres can be transferred directly to the case of 
inelastic hard spheres.
This will be done by deriving the Boltzmann equation and the ring kinetic 
equation for the model under consideration.

The Boltzmann equation is obtained from the first hierarchy
equation by keeping only terms to dominant order in the density.
This implies the
fundamental assumption of molecular chaos for dilute gases, 
expressing the absence of 
dynamic correlations between the velocities of two colliding particles 
just {\em before} collision, i.e. at $|{\bf r}_{12}|=\sigma+0$,
\be 
f_{12}=f({\bf r}_1,{\bf v}_1,t) f({\bf r}_2,{\bf v}_2,t).
\label{eq:molchaos}
\ee
Furthermore, in the low density limit the spatial separation
between
the colliding particles can be neglected, and the binary collision
operator $\overline{T}(12)$, entering in the BBGKY hierarchy, reduces
to
\be
\overline{T}(12)=\delta({\bf r}_{12})\overline{T}_0(12)=\delta(
{\bf r}_{12})\,\sigma^{d-1}\int_{{\bf v}_{12}\cdot\hat{\sigma}>0}
{\rm d}\hat{\bsigma} ({\bf
v}_{12}\cdot\hat{\bsigma})\left(\frac{1}{\alpha^2}
b^{\ast\ast}_{\sigma}-1\right).
\ee
Then the nonlinear Boltzmann equation for inelastic hard spheres becomes
\ba
&&(\partial_t+ L_1^0) f({\bf r}_1,{\bf v}_1,t) =
\sigma ^{d-1} \int {\rm d}{\bf v}_2 \int_{{\bf v}_{12}\cdot\hat{\sigma}>0} 
{\rm d}\hat{\bsigma} ({\bf v}_{12}\cdot\hat{\bsigma})\times\nonumber\\
&&\left\{\frac{1}{\alpha^2}f({\bf r}_1,{\bf v}_1^{\ast\ast},t)f({\bf r}_1,
{\bf v}_2^{\ast\ast},t)-f({\bf r}_1,{\bf v}_1,t)f({\bf r}_1,{\bf v}_2,t)
\right\}\equiv I(f,f).
\label{eq:boltz}
\ea
There are two significant differences with the Boltzmann equation
for the elastic case: (i) the occurrence of $1/\alpha^2$ in
the gain term on the right hand side of (\ref{eq:boltz}); one
factor $1/\alpha$ comes from the Jacobian ${\rm d}{\bf
v}_1^{\ast\ast} {\rm d}{\bf
v}_2^{\ast\ast}=(1/\alpha) {\rm d}{\bf
v}_1 {\rm d}{\bf v}_2$ and the other one from the reflection law
${\bf v}_{12}^{\ast\ast}\cdot\hat{\bsigma}=-(1/\alpha){\bf
v}_{12}\cdot\hat{\bsigma}$ (see Fig.\ 1).
(ii) In the inelastic case, the {\em restituting} precollision
velocities, which yield $({\bf v}_1,{\bf v}_2)$ as postcollision
velocities, are {\em different} from the postcollision velocities
$({\bf v}_1^\ast,{\bf v}_2^\ast)$, which result from the {\em
direct} precollision velocities.
In the elastic case $(\alpha=1)$ the relation ${\bf v}_i^\ast={\bf
v}_i^{\ast\ast}$ holds.

The Boltzmann-Enskog equation for inelastic hard spheres (see
Refs.\
\cite{lun,jenkins+richman,brey+dufty+santos,goldshtein}) 
is obtained by
replacing $f_{12}$ in the first hierarchy equation of
(\ref{eq:bbgky})
by $\chi({\bf r}_1,{\bf r}_2|n)f_1f_2$, where $\chi({\bf r}_1,{\bf
r}_2|n)$ is the pair correlation function of elastic hard spheres in
a spatially
nonuniform equilibrium state (see Ref. \cite{vbeijeren+ernst}).
This version of the molecular chaos assumption still neglects the
velocity correlations, built up by sequences of correlated binary
collisions, but does account for static short range correlations,
caused by excluded volume effects.
The consequence for the transport coefficients have been worked out
in
Refs.\
\cite{lun,jenkins+richman,goldshtein}.
 
As the density increases the contributions of correlated collision
sequences to the collision term on the right hand side of 
(\ref{eq:enskog}) become more and more important.
The most simple sequence of correlated collisions
are the so called {\em ring} collisions; for example
(12)(13)(14)$\dots$(23$'$)(24$'$)$\dots$(12), ending with a 
recollision of
the pair (12), which was involved in the first collision. 
In the intermediate time
particle 1 collides, say, $s$ times with $s$ different particles 
(3,4,$\dots$), and particle
2 collides $s'$ times with {\em another} set of $s'$ different particles 
(3$'$,4$'$,$\dots$). 
When particles 1 and 2 are about to recollide, they are dynamically 
correlated
through their collision history, and the molecular chaos assumption
(\ref{eq:molchaos}) is no longer valid, i.e. $g_{12}\equiv f_{12}-
f_1 f_2\neq 0$.
 
A simple way to take these correlations into account at moderate
densities has been given in Refs.\ 
\cite{cluster,dorfman+cohen-cluster}. 
The method is based on a cluster expansion of the $s$-particle
distribution functions, defined recursively as
\ba
f_{12}&=&f_1 f_2 + g_{12}\nonumber\\
f_{123}&=&f_1 f_2 f_3 +f_1 g_{23} + f_2 g_{13} + f_3 g_{12} + g_{123},
\label{eq:cluster}
\ea
etc.
Here $g_{12}$ accounts for pair correlations, $g_{123}$ for triplet
correlations, etc. 
The molecular chaos assumption implies $g_{12}=0$, which is
equivalent to (\ref{eq:molchaos}).
The basic assumption to obtain the ring kinetic
equation is that the pair correlations are dominant and higher order
ones can be neglected, i.e. $g_{123}=g_{1234}=\cdots=0$ in cluster
expansion (\ref{eq:cluster}).
 
Substitution of (\ref{eq:cluster}) into (\ref{eq:bbgky}) and elimination 
of $\partial f_i/\partial t \,\, (i=1,2)$
from the second hierarchy equation using the first one, yields the
ring kinetic theory of inelastic hard spheres:
\ba
&&(\partial_t  +  L_1^0) f_1 = \int {\rm d} x_2 
\overline{T}(12) (f_1 f_2 +g_{12})
\label{eq:firstring}\\
&&[\partial_t +  L_1^0  + L_2^0 - \overline{T}(12) - 
(1+{\cal P}_{12}) \int {\rm d} x_3 \overline{T}(13) (1+{\cal P}_{13}) f_3 ]
g_{12} = \overline{T}(12) \left[f_1 f_2+g_{12}\right].
\label{eq:secondring}
\ea
Here ${\cal P}_{ij}$ is a permutation operator that interchanges the 
particle labels $i$ and $j$.
The second equation is the so called {\em repeated ring} equation for
the pair correlation function. 
If the operator ${\overline T}(12)$ on
the left hand side of the second equation is deleted, one obtains the
simple {\em ring} approximation. 
Formally solving this equation for $g_{12}$ yields
an expression in terms of the single-particle distribution functions
$f_i\,\, (i=1,2,3)$, and subsequent substitution into the first
hierarchy equation above yields the generalized Boltzmann
equation in ring approximation. For a more detailed discussion of the
collision sequences taken into account by Eq.\ (\ref{eq:secondring}) 
we refer to the original literature 
\cite{cluster,dorfman+cohen-cluster,sengers}.

The kinetic equations (\ref{eq:firstring}) and (\ref{eq:secondring}) 
constitute the new extensions of this article: the ring kinetic
equations for inelastic hard spheres.
All standard results for elastic hard spheres are recovered by setting the 
restitution coefficient $\alpha=1$. 
In this paper we will determine from (\ref{eq:secondring}) the
behavior of the nonequilibrium pair distribution function
$g(x_1,x_2,t)$ on hydrodynamic time scales, for a special solution
$f(x,t)$ of the Boltzmann equation.
The spatial correlations $\langle u_\alpha(\br+\br^\prime,t)
u_\beta(\br^\prime,t)\rangle$ of the flow field ${\bf u}(\br,t)$,
calculation of which is our direct goal in this paper, can be
directly obtained from $g(x_1,x_2,t)$ by integration.
In the next section will summarize some known
results about the homogeneous cooling state which are necessary
for an understanding of what follows later on.

\section{Homogeneous Cooling State}
\label{sec:HCS}
A gas of {\em elastic} hard spheres will relax to local equilibrium on a 
(kinetic) time 
scale of a few mean free times, $t_0\sim l_0/v_0$, where $l_0$ is the 
mean free path and $v_0$ the thermal velocity, $\textstyle{\frac{1}
{2}}m v_0^2\equiv T$.
Here the Boltzmann constant $k_B$ is set equal to 
unity.
Finally, the system will reach a spatially homogeneous equilibrium state 
on a (hydrodynamic) time scale $\sim L/v_0$, where $L$ is the linear dimension of the
system.

However, in the case of {\em inelastic} hard spheres kinetic energy is
lost in collisions,
and if the system is not driven, the kinetic energy associated with the 
thermal motion decreases, and interesting
instabilities occur, such as clustering \cite{goldhirsch}
and inelastic collapse \cite{mcnamara}. 
 
Here we are interested in a special solution of the Boltzmann equation 
(\ref{eq:boltz}) for inelastic hard spheres, the so called 
{\em homogeneous cooling state} (HCS).
In this state the distribution function $f({\bf r},{\bf v},t)=
f({\bf v},t)$, as well as the hydrodynamic fields are spatially
{\em homogeneous}.
These functions are defined as density 
$n({\bf r},t)=\int {\rm d}{\bf v} f({\bf r},{\bf v},t)$, flow velocity 
${\bf u}({\bf r},t)=\left(1/n({\bf r},t)\right)\int {\rm d}{\bf v} {\bf v} 
f({\bf r},{\bf v},t)$, and {\em granular temperature} $T({\bf r},t)=(m/d
n({\bf r},t))\int
{\rm d}{\bf v} V^2 f({\bf r},{\bf
v},t)$,
where ${\bf V}({\bf r},t)={\bf v}-{\bf u}({\bf r},t)$ is the peculiar 
velocity. 
Furthermore, the flow velocity can be taken to vanish, ${\bf u}({\bf r},
t)={\bf 0}$, and $n({\bf r},t)=n$ is constant in space and time.
However, $T({\bf r},t)=T(t)$ depends on time.
Based on the fundamental concepts of the Chapman-Enskog theory, 
we expect that the single-particle distribution function $f$ after an 
initial transient of the order of a few mean free times, will only 
depend on time through  
its first few moments, which is here only the temperature $T(t)$.
For dimensional reasons $f({\bf v},t)$ then takes the scaling form
\be
f({\bf v},t)=\frac{n}{v_0^d(t)}\tilde{f}\left(\frac{\bf v}{v_0(t)}\right),
\label{eq:scaling}
\ee
with the thermal velocity $v_0(t)$ depending on time.
One can derive an integral equation for the unknown scaling form 
$\tilde{f}({\bf c})$, with ${\bf c}={\bf v}/v_0(t)$, by inserting 
(\ref{eq:scaling}) into the Boltzmann equation (\ref{eq:boltz}).
The result is
\be
-\frac{1}{v_0^2}\frac{d v_0}{d t}\left(d +{\bf c}_1\cdot\frac{
{\rm d}}{{\rm d} {\bf c}_1} \right) \tilde{f}({\bf c}_1)=
n\sigma^{d-1} \widetilde{I}(\tilde{f},
\tilde{f}),
\label{eq:inteq1}
\ee
with
\be
\widetilde{I}(\tilde{f},
\tilde{f})\equiv \int {\rm d} {\bf c}_2 \int_{{\bf c}_{12}
\cdot\hat\sigma>0} {\rm d} \hat{\bsigma} ({\bf c}_{12}\cdot\hat{\bsigma}) 
\left(
\frac{1}{\alpha^2} b^{\ast\ast}_{\sigma}-1\right) \tilde{f}({\bf c}_1)
\tilde{f}({\bf c}_2).
\ee
To determine the rate of change of the temperature, we use Eq.\ 
(\ref{eq:ratepsi}) with $\psi({\bf v})= v^2$
and $f_{12}=f({\bf v}_1,t) f({\bf v}_2,t)$, and calculate 
${v_1^\ast}^2-v_1^2$.
With the help of Eqs.\ (\ref{eq:collr}) and 
(\ref{eq:scaling}), we obtain
\be
\frac{{\rm d} T}{{\rm d} t} = - \frac{\Omega_d}{\sqrt{2\pi}}
m n \sigma^{d-1} v_0^3 \gamma= -2 \omega \gamma T,
\label{eq:temprate}
\ee
where $\omega$ is the time dependent collision frequency given in
Boltzmann theory by
$\omega=\Omega_d n \sigma^{d-1}\sqrt{T/\pi m}$ with $\Omega_d=2
\pi^{d/2}/\Gamma(d/2)$ the
surface area of a $d$-dimensional unit sphere, and $\gamma$ 
is a time independent cooling rate, defined by
\ba
\gamma &\equiv& -\frac{\sqrt{2\pi}}{d\Omega_d} \int {\rm d}{\bf c}_1 c_1^2 
\widetilde{I}
(\tilde{f},\tilde{f})\nonumber\\
&=& \left(\frac{\sqrt{2\pi}}{d\Omega_d}\right)\frac{1-\alpha^2}{4}
\int {\rm d}{\bf c}_1\int 
{\rm d}{\bf c}_2 
\int_{{\bf c}_{12}\cdot\hat\sigma>0} {\rm d}\hat{\bsigma} 
({\bf c}_{12}\cdot \hat{\bsigma})^3 \tilde{f}({\bf c}_1)\tilde{f}
({\bf c}_2),
\label{eq:gamma}
\ea
which depends on the unknown scaling form $\tilde{f}({\bf c})$.

At this stage, it is convenient to change to a new time variable
$\tau$,
defined through ${\rm d}\tau=\omega(T(t)) {\rm d} t$.
Integration yields
\be
\tau=\frac{1}{\gamma}\ln(1+\gamma t/t_0).
\label{eq:tau}
\ee
The time $\tau$ actually presents the average number of collisions
suffered per particle.
In the elastic limit ($\eps\rightarrow 0$), it becomes the real
time,
$\tau=t/t_0$, measured in units of the mean free time
$t_0=1/\omega(T_0)$ at the initial
temperature $T_0$.

Eq.\ (\ref{eq:temprate}) is readily integrated to obtain for the 
temperature
\be
T(t)=\frac{T_0}{(1+ \gamma t/t_0)^2}=T_0\exp(-2 \gamma \tau).
\label{eq:v0decay}
\ee
The above equation represents the well known algebraic decay law
for the granular temperature in the homogeneous cooling state\footnote{
Note that, because of the spatial homogeneity of $f$,
this result is still valid in Enskog theory, for general 
densities, once the collision frequency has been
increased by the factor $\chi(n)$, which is the local equilibrium 
pair correlation
function at contact.}
(see Refs.\ \cite{goldhirsch,mcnamara} and references therein).

Combining (\ref{eq:inteq1}) and (\ref{eq:temprate}) yields an integral 
equation for the scaling form $\tilde{f}({\bf c})$, i.e.\
\be
\frac{\Omega_d}{\sqrt{2\pi}}\gamma \left(d+{\bf c}_1\cdot\frac{{\rm
d}}{{\rm d}
{\bf c}_1}\right) 
\tilde{f}({\bf c}_1) = \widetilde{I}(\tilde{f},\tilde{f}),
\label{eq:inteq2}
\ee
where $\gamma$ is given through (\ref{eq:gamma}) in terms of $\tilde{f}$ 
itself.
By a moment expansion of this equation, it has been shown 
\cite{noije+brito+ernst} that
the fourth cumulant of $\tilde{f}$ is small for any value of the 
inelasticity $\eps$
both in $d=2$ and 3.
This has been confirmed by a direct simulation Monte Carlo calculation
of the fourth and sixth cumulant in the homogeneous cooling state
of inelastic hard spheres ($d=3$) \cite{brey+montanero+cubero}.
To a good approximation, therefore, the solution of
(\ref{eq:inteq2}) approaches a Maxwellian, i.e.\
$\tilde{f}(c)\approx \phi(c) \equiv \pi^{-d/2} \exp(-c^2)$.
Similarly, the dimensionless cooling rate (\ref{eq:gamma}) is well
approximated by replacing $\tilde{f}(c)$ by $\phi(c)$ in
(\ref{eq:gamma}):
\be
\gamma_0=-\frac{\sqrt{2\pi}}{d\Omega_d} \int {\rm d}{\bf c}_1 c^2_1 \widetilde{I}(\phi,\phi)=\frac{1-\alpha^2}{2d}.
\label{eq:gamma0}
\ee

\section{Ring kinetic theory}
\label{sec:corrtheory}
In this section we will derive the structure factor of transverse
velocity or vorticity
fluctuations, $S_\perp(k,t)$, for a freely evolving, dilute gas of 
inelastic hard spheres.
In general velocity correlations can be described by an isotropic
tensor $S_{\alpha\beta}(\bk,t)$, related to the 
scalar isotropic functions $S_\perp(k,t)$ and $S_\parallel(k,t)$ 
via the decomposition \cite{batchelor}
\be
S_{\alpha\beta}({\bf k},t)=\hat{k}_\alpha\hat{k}_\beta
S_\parallel(k,t)+(\delta_{\alpha\beta}-\hat{k}_\alpha\hat{k}_\beta)
S_\perp(k,t),
\ee
where hats denote unit vectors.
The structure factors $S_\perp(k,t)$ and $S_\parallel(k,t)$ of
transverse and longitudinal velocity fluctuations are also related 
to the energy spectrum function $E(k)$ in the theory of
homogeneous turbulence \cite{batchelor}.

The inverse Fourier transform of $S_{\alpha\beta}(\bk,t)$ are
the spatial correlations
$G_{\alpha\beta}(\br,t)$.
In terms of the
microscopic velocity field ${\bf u}(\br,t)$, it may be defined as
\be
G_{\alpha\beta}(\br,t)=\frac{1}{V} \int {\rm d}{\bf r}^\prime \langle
u_\alpha({\bf r}+\br^\prime,t) u_\beta({\bf r}^\prime,t)\rangle.
\ee
A similar decomposition \cite{batchelor} holds for the spatial
velocity correlations:
\be
G_{\alpha\beta}(\br,t)=\hat{r}_\alpha\hat{r}_\beta
G_\parallel(r,t)+(\delta_{\alpha\beta}-\hat{r}_\alpha
\hat{r}_\beta)
G_\perp(r,t).
\ee
Whereas the tensors $G_{\alpha\beta}(\br,t)$ and 
$S_{\alpha\beta}(\bk,t)$
include self-correlations of particles, the correlation
functions $g_{12}$ and $s_{12}$ (as defined below) occurring in the 
second hierarchy equation do not.
Therefore, it is convenient to substract the self-correlation part
and 
introduce the tensors
$S^+_{\alpha\beta}(\bk,t)\equiv S_{\alpha\beta}(\bk,t)-T(t)
\delta_{\alpha\beta}/n m$ and $G^+_{\alpha\beta}(\br,t)\equiv 
G_{\alpha\beta}(\br,t)
-T(t)\delta_{\alpha\beta} \delta(\br)/ n m$, which is regular at
the origin and related to $g_{12}$ by
\be
G^+_{\alpha\beta}(\br,t)=\frac{1}{V}\int {\rm d}\br^\prime
\frac{1}{n^2} \int {\rm d}{\bf v}_1 \int {\rm d}{\bf v}_2
v_{1\alpha}v_{2\beta} g(\br+\br^\prime,{\bf v}_1,\br^\prime,{\bf
v}_2,t).
\ee
Incompressibility of velocity fluctuations then implies
$S^+_\parallel(k,t)=0$, in which case
$G^+_\perp(r,t)$ is related to $G^+_\parallel(r,t)$ by (see
\cite{batchelor})
\be
G^+_\perp(r,t)=G_\parallel^+(r,t)+\frac{r}{d-1}
\frac{\partial}{\partial r} G_\parallel^+(r,t).
\ee

As discussed in section \ref{sec:BBGKY}, in the low density limit,
$\overline{T}(12)=\delta({\bf r}_{12}) \overline{T}_0(12)$,
the first hierarchy equation reduces to the Boltzmann equation,
and the second hierarchy equation to its simple ring
approximation,
\be
\left(\frac{\partial}{\partial t} + {\bf v}_1\cdot\frac{\partial}
{\partial {\bf r}_1} +\Omega_1 + 
{\bf v}_2\cdot\frac{\partial}{\partial {\bf r}_2} +\Omega_2\right) g_{12} 
= \delta({\bf r}_{12})
\overline{T}_0(12) f_1 f_2, 
\label{eq:second}
\ee
where $\Omega$ is minus the linearized Boltzmann collision operator,
i.e.\ ($i=1,2$)
\ba
\Omega_i \psi({\bf v}_i) &\equiv& -\int {\rm d} x_3
\overline{T}_0(i3)(1+{\cal P}_{i3}) f_3 \psi({\bf v}_i)\nonumber\\
&=&-\sigma^{d-1} \int {\rm d}{\bf v}_3 \int_{{\bf
v}_{i3}\cdot\hat{\sigma}>0} {\rm d}\hat{\bsigma} ({\bf
v}_{i3}\cdot\hat{\bsigma})\left(\frac{1}{\alpha^2}
b^{\ast\ast}_{\sigma}-1\right) \times\nonumber\\
&&\left\{ f({\bf v}_i,t)\psi({\bf v}_i)+ f({\bf v}_3,t)\psi({\bf
v}_3)\right\}.
\ea

The term on the right hand side of Eq.\ (\ref{eq:second}), which vanishes 
in the case of detailed balance, provides a source of correlations for
inelastic hard spheres in the homogeneous cooling state.
For a homogeneous distribution $f({\bf v},t)$, $g(\br_1,{\bf
v}_1,\br_2,{\bf v}_2,t)$ depends on $\br_{12}=\br_1-\br_2$ only,
and
the ring equation 
for the 
Fourier transform
$s(\bk,{\bf v}_1,{\bf v}_2,t)\equiv\int {\rm d}{\br}_{12} e^{-i \bk\cdot
\br_{12}} g(\br_1,{\bf v}_1,\br_2,{\bf v}_2,t)$ is given by 
\be
(\partial_t + i\bk\cdot {\bf v}_{12} +\Omega_1+\Omega_2) s_{12}=
\overline{T}_0(12) f({\bf v}_1,t) f({\bf v}_2,t).
\label{eq:second2}
\ee
Inspired by the scaling ansatz (\ref{eq:scaling}) for the homogeneous 
cooling solution of 
the Boltzmann equation, we write the density and temperature dependence 
of $s_{12}$ explicitly as 
\be
s({\bf k},{\bf v}_1,{\bf v}_2,t)=\frac{n^2}{v_0^{2d}(t)} \tilde{s}({\bf k},{\bf c}_1,{\bf c}_2,\tau).
\ee
Again it is convenient to transform to the kinetic time $\tau$, so
that
Eq.\ (\ref{eq:second2}) reduces to
\be
[\partial_\tau + \widetilde{\Lambda}_{12}({\bf k})] \tilde{s}
(\bk,{\bf c}_1,{\bf c}_2,\tau)= l_0 \sigma^{d-1} \widetilde{T}_0(12)
\tilde{f}({\bf c}_1)\tilde{f}({\bf c}_2),
\ee
with the mean free path $l_0=v_0/\omega$.
Here we have introduced the notation ($j=1,2$)
\ba
\widetilde{\Lambda}_j({\bf k})&=& i l_0 {\bf k}\cdot{\bf c}_j
+\gamma\left(d+
{\bf c}_j\cdot \frac{\partial}{\partial{\bf c}_j}\right) +
l_0 n \sigma^{d-1} \tilde{\Omega}_j
\nonumber\\
\widetilde{\Lambda}_{12}({\bf k})&=&\widetilde{\Lambda}_1({\bf k})+
\widetilde{\Lambda}_2(-{\bf k}),
\ea
defined in terms of the dimensionless quantities
\ba
\widetilde{\Omega}&\equiv&\frac{1}{v_0(t)\sigma^{d-1}n} \Omega_1
\nonumber\\
\widetilde{T}_0(12)&=&\frac{1}{v_0(t) \sigma^{d-1}}\overline{T}_0(12).
\ea

The formal solution for the case that all pair correlations vanish at the 
initial time $t=0$, is given by
\ba
\tilde{s}(\bk,{\bf c}_1,{\bf c}_2,\tau)&=& \frac{1-\exp[-\tau 
\widetilde{\Lambda}_{12}(\bf k)]}{\widetilde{\Lambda}_{12}(\bf k)} l_0 
\sigma^{d-1} \widetilde{T}_0(12) \tilde{f}({\bf c}_1)\tilde{f}({\bf c}_2)
\nonumber\\
&=&\sum_{\lambda,\mu} -\frac{\exp{\tau[z_\lambda(\bk)+
z_\mu(\bk)]}-1}{z_\lambda(\bk)+z_\mu(\bk)}
|\psi^R_\lambda({\bf c}_1,{\bf k}) \psi^R_\mu({\bf c}_2,{\bf k}) \rangle 
\times\nonumber\\
&&\langle \psi^L_\lambda({\bf c}_1,{\bf k}) \psi^L_\mu({\bf c}_2,{\bf k})
| l_0 \sigma^{d-1} \widetilde{T}_0(12) \tilde{f}({\bf c}_1)\tilde{f}
({\bf c}_2)\rangle.
\label{eq:composition}
\ea
In the second line we have made a decomposition in eigenfunctions 
of the operator $\widetilde{\Lambda}$ and used the bracket notation,
with inner products representing integrals over ${\bf c}$. 
The left and right eigenfunctions satisfy the eigenvalue
relations,
\ba
\widetilde{\Lambda}_j({\bf k})|\psi^R({\bf c}_j,{\bf k})\rangle &=& 
-z_\lambda({\bf k}) |\psi^R({\bf c}_j,{\bf k})\rangle\nonumber\\
\langle \psi^L({\bf c}_j,{\bf k})| \widetilde{\Lambda}_j({\bf k})&=& 
-z_\lambda ({\bf k}) \langle \psi^L({\bf c}_j,{\bf k})|.
\ea

The structure factor of transverse velocity fluctuations,
$S^+_\perp(k,t)$, is then given by
\ba
S^+_\perp(k,t)&=&\frac{1}{n^2}\int {\rm d}{\bf v}_1 \int {\rm d}{\bf v}_2
v_{1\perp\alpha} v_{2\perp\alpha}
s(\bk,{\bf v}_1,{\bf v}_2,t)\nonumber\\
&=&v_0^2(t)\langle c_{1\perp\alpha} 
c_{2\perp\alpha} | \tilde{s}(\bk,{\bf c}_1,
{\bf c}_2,\tau)\rangle.
\ea
We state here without further derivation that the transverse
velocity or shear modes are the only modes contributing to
$S^+_\perp(k,t)$ (see also \cite{boston,noije+ernst+brito+orza}).
It can be shown that the eigenfunctions corresponding to the shear
mode are $\psi_{\perp\alpha}={\bf c}\cdot\hat{\bf
k}_{\perp\alpha}=
c_{\perp\alpha}$, where the subscript $\alpha$ denotes one of
the $d-1$ degenerate shear modes.
The dispersion relation for the relaxation rate $z_\perp(k)$ of the
shear mode with wavenumber $k$ is $z_\perp(k)=\gamma_0(1-k^2
\xi^2)$.
The correlation length $\xi=\sqrt{\eta/\rho \omega\gamma_0}$ with
$\eta\sim \sqrt{T(t)}$ the time dependent shear viscosity, $\rho=m
n$ the mass density, $\omega\sim \sqrt{T(t)}$ the collision
frequency and $\gamma_0=\eps/2 d$.
For small inelasticity $\xi$ diverges as $1/\sqrt{\eps}$.
So in calculating $S^+_\perp(k,t)$, we only have to calculate
the quantity $\langle c_{1\perp\alpha} 
c_{2\perp\alpha} | l_0 \sigma^{d-1} \widetilde{T}_0
(12) \tilde{f}({\bf c}_1)\tilde{f}({\bf c}_2)\rangle= -\Omega_d
l_0 \sigma^{d-1} \gamma/\sqrt{2\pi}=-\gamma/n\approx -\gamma_0/n$, 
as can easily be shown
using Eq.\ (\ref{eq:ratepsi}).
We then obtain for $S^+_\perp(k,t)$ the expression
\be
S^+_\perp(k,t)= \left(\frac{T(t)}{n m}\right)\frac{
\exp[2 \gamma_0 \tau(1-k^2\xi^2)]-1}{1-k^2 \xi^2}.
\label{eq:transverse}
\ee
This low density result agrees with a previous result derived from
fluctuating hydrodynamics \cite{noije+ernst+brito+orza},
which, moreover, extends the validity of the above expression to
higher densities.

The subsequent analysis of $G_\perp(r,t)$ and $G_\parallel(r,t)$,
under the simplifying assumption that the fluctuating flow fields
are {\em incompressible}, i.e.\ $S_\parallel^+(k,t)=0$, has been
given in \cite{noije+ernst+brito+orza}.
The full analysis for the compressible case can also be given, and
yields only small modifications to the incompressible case, except
at the largest scales, where the algebraic tails are cut off
exponentially \cite{noije+brito+ernst}.

It is found that the spatial correlation functions $G_\perp(r,t)$
and $G_\parallel(r,t)$ show structure on spatial scales large
compared to the mean free path $l_0$.
In particular one obtains for asymptotically large $r$ the
algebraic tails, $G_\parallel(r,t)\sim
-(d-1)
G_\perp(r,t)\sim A/r^{d}$ with explicit expressions for $A$ 
\cite{noije+brito+ernst}.
Furthermore, $G_\parallel(r,t)$ is positive everywhere, while
$G_\perp(r,t)$ has a negative minimum, reflecting the presence of
vortices in the system.
The structure function $S^+_\perp(k,t)$ for the flow field in the
inelastic hard sphere system is the analog of the energy spectrum
function $E(k)$ in the theory of 2-$D$ or 3-$D$ homogeneous turbulence 
in incompressible fluids. 
The qualitative forms of $G_\perp(r,t)$ and $G_\parallel(r,t)$ are
roughly similar to the shapes of these functions in the theory of
homogeneous turbulence, as illustrated in Fig.\ 3.2 of chapter 3 in
Ref.\ \cite{batchelor}.
For a more detailed description of these functions, we refer to
\cite{noije+brito+ernst}.

We now return to the structure factors, and compare Eq.\
(\ref{eq:transverse}) with results from a two-dimensional molecular 
dynamics
simulation of $N=50000$ inelastic hard disks at a low area fraction 
$\phi=\textstyle{\frac{1}{4}}\pi
n \sigma^2=0.05$, and a coefficient of normal restitution $\alpha=0.85$. 
Fig.\ 2a shows the prediction of Eq.\ (\ref{eq:transverse}) in the
low density limit at $\tau=30$ collisions per particle, together
with simulation results for $S_\perp(k,t)$ and $S_\parallel(k,t)$, 
as obtained
by performing a Fourier transform on the
momentum fields, coarse-grained
into $256\times 256$ boxes, followed by a circular average.
At the corresponding time $t$,  
Eq.\ (\ref{eq:tau}),
gives the slightly smaller
value $\tau=28.7$ for the number of collisions per particles.
Already at this density, Enskog theory gives a quantitatively
significant increase
of the collision frequency by a factor $\chi\simeq 1.08$, 
leading to a predicted number of collisions per particle $\tau=29.7$,
for the case under consideration.
For a more detailed comparison of both $S_\perp(k,t)$ and
$S_\parallel(k,t)$ at general densities, we refer to
\cite{noije+brito+ernst}.
Since the structure factor of transverse velocity fluctuations, 
$S^+_\perp(k,t)$,
shows structure at wavenumbers $k\lesssim 1/\xi$, $\simeq 0.06\sigma^{-1}$ in the ring kinetic theory,
the spatial velocity correlation functions 
$G_\perp(r,t)$ and $G_\parallel(r,t)$ (shown in Fig.\ 2b) 
also show structure up to and beyond
distances of the order $2 \pi \xi\simeq 106\sigma$, which is large
compared to the Boltzmann mean free path $l_0\simeq 6.3\sigma$,
but still small compared to the system size $L=886\sigma$.
Furthermore, the simulation results show that the incompressibility
assumption indeed holds in a range of wavenumbers
$1/\xi_\parallel \lesssim k <1/l_0$, 
where the existence of a large distance cut-off length
$2\pi \xi_\parallel \gg 2\pi \xi$ has been discussed in
Ref.\ \cite{noije+brito+ernst}.
The spatial velocity correlations measured in the simulations agree
well with the calculated $G_\perp(r,t)$ and $G_\parallel(r,t)$ in
the range $2\pi l_0<r\lesssim 2\pi \xi_\parallel$, 
and exhibit an observable
long range $r^{-d}$-tail for distances $2\pi \xi
\lesssim r \lesssim 2 \pi \xi_\parallel$.
At wavenumbers of the order of the minimal accessible wavenumber $k_{\rm
min}=2\pi/L\simeq 0.007\sigma^{-1}$, effects from the periodic
boundaries become 
important.
To conclude, we observe that the more fundamental ring kinetic
theory yields results identical to the more mesoscopic theory of
fluctuating hydrodynamics in their common region of validity, thus
supporting the phenomenological theory presented in Ref.\
\cite{noije+ernst+brito+orza,noije+brito+ernst}.

\\
\\
The authors want to thank R. Brito and 
J.A.G. Orza, who performed the molecular dynamics simulations, for
fruitful collaboration, and
H.J. Bussemaker and D.
Montgomery for stimulating discussion.
T.v.N. acknowledges support of the
foundation `Fundamenteel Onderzoek der Materie (FOM)', which is
financially supported by the Dutch National Science Foundation
(NWO).

\begin{figure}
\centerline{\psfig{file=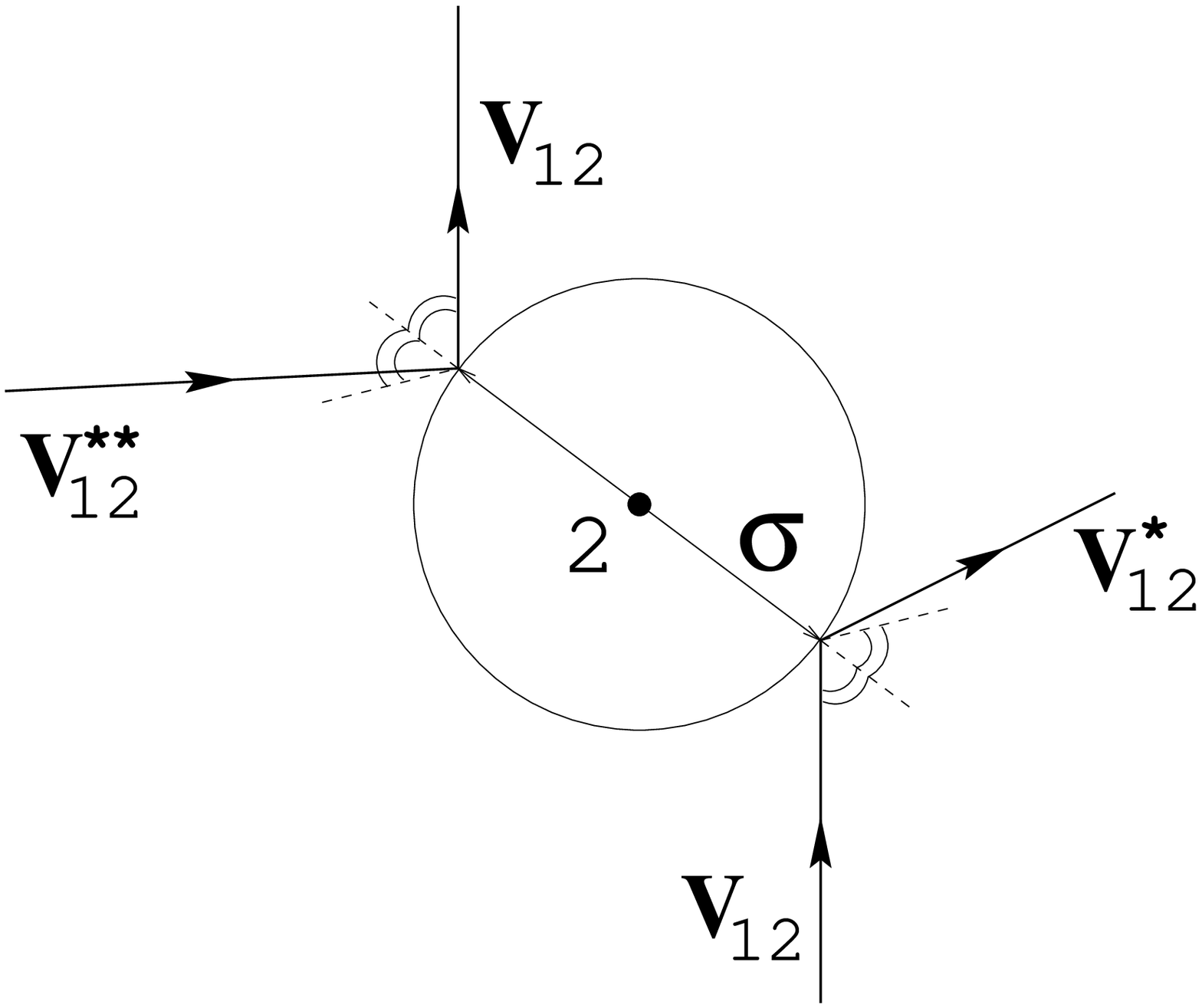,height=6cm}}
\vspace*{1cm}
\caption{Direct and restituting inelastic collisions with
refection law ${\bf v}_{12}^\ast\cdot\hat{\bsigma}=-\alpha {\bf
v}_{12}\cdot\hat{\bsigma}$ with $0<\alpha<1$.}
\end{figure}
\begin{figure}
\hspace*{1cm}
\parbox[b]{7cm}{
\centerline{\psfig{file=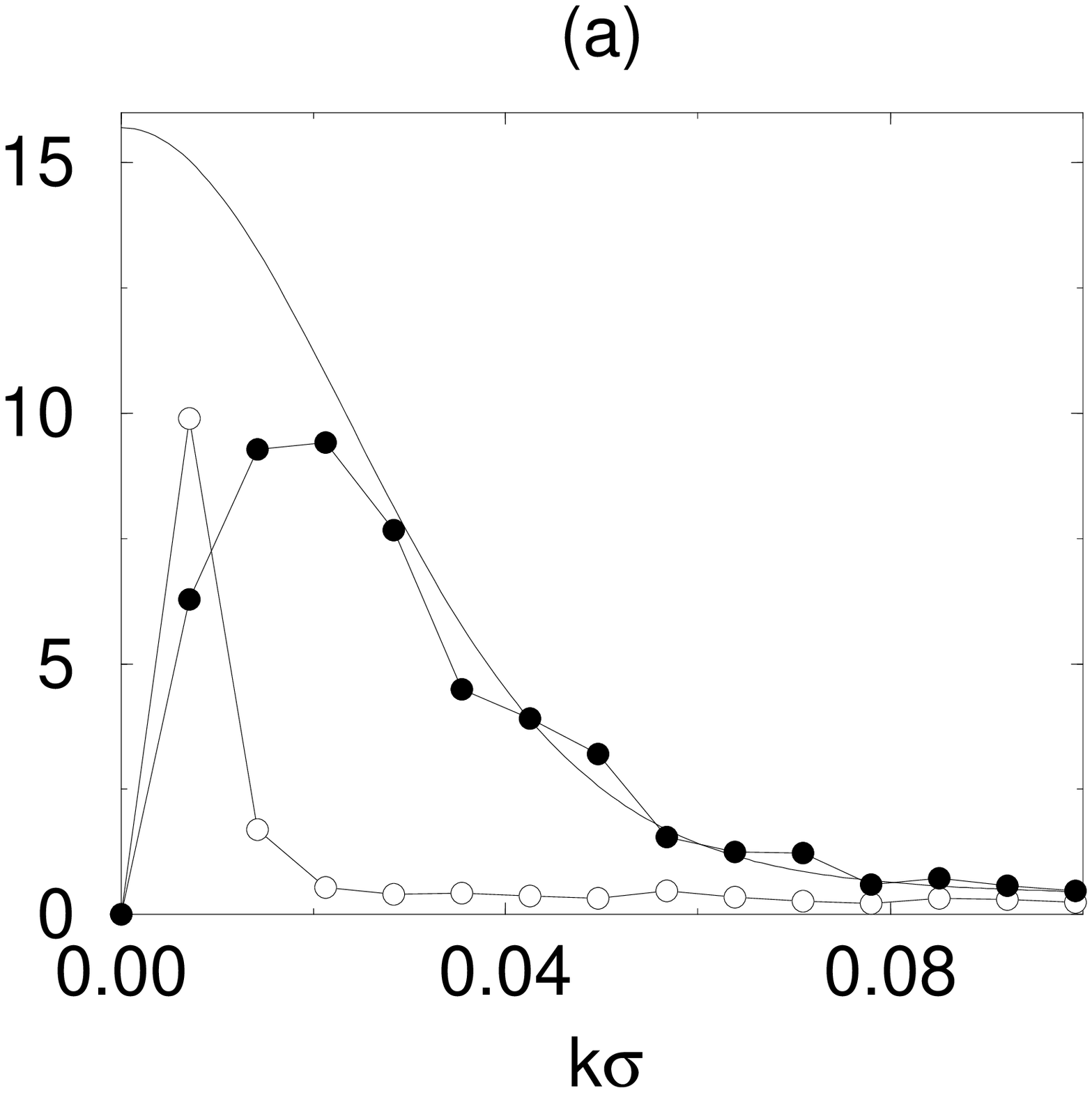,height=6cm}}
}
\parbox[b]{7cm}{
\centerline{\psfig{file=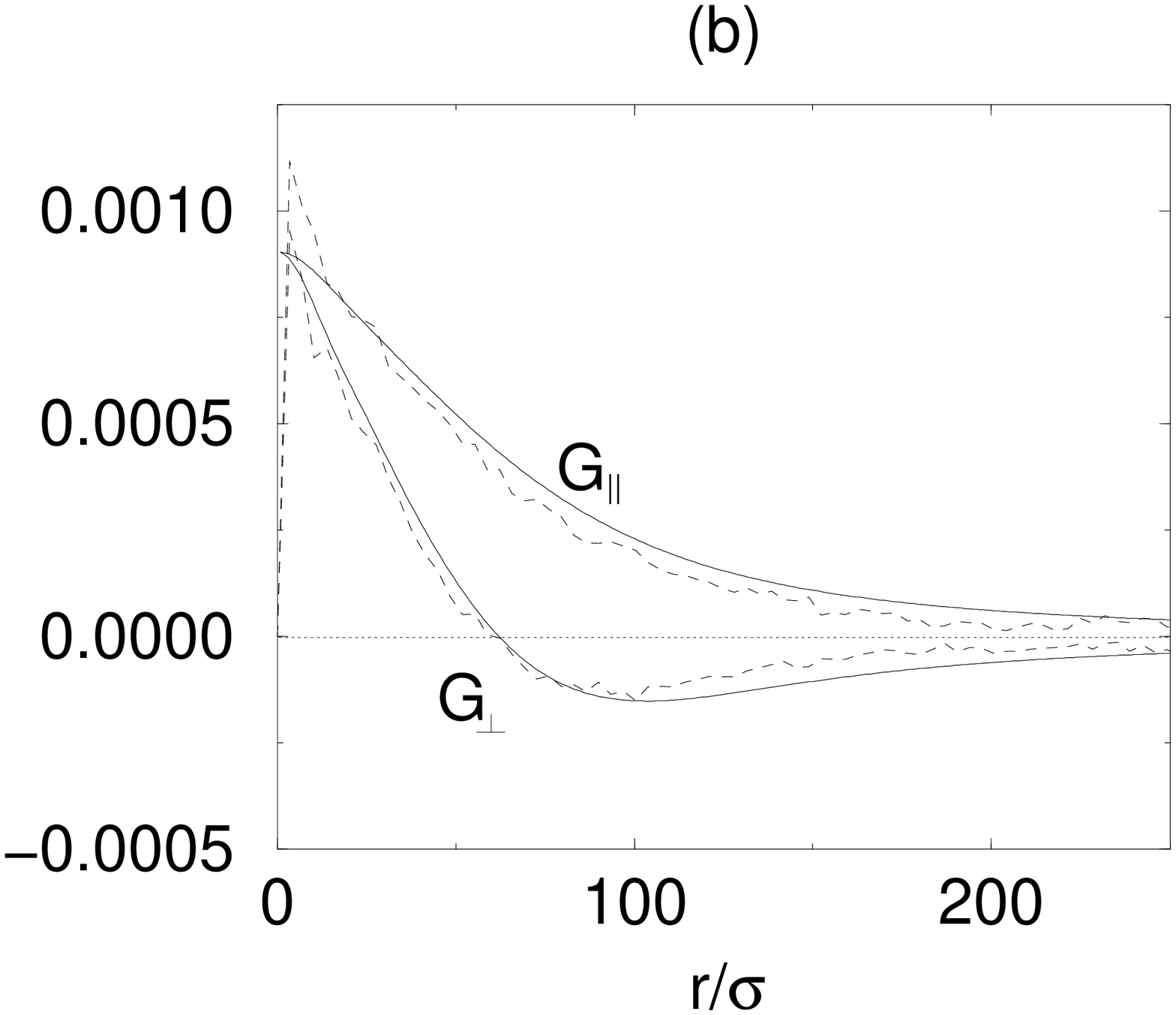,height=6cm}}
}
\vspace*{1cm}
\caption{
(a) $S_\perp(k,t)$ (filled circles) and $S_\parallel(k,t)$
(open
circles) as
measured from molecular dynamics simulations of $N=50000$
particles at an area fraction
$\phi=0.05$, coefficient of normal restitution
$\alpha=0.85$ and
number of collisions per particle $\tau=30$, with $T_0/m=1$,
compared with the prediction
Eq.\ (\ref{eq:transverse}) from ring kinetic theory for
$S_\perp(k,t)$ (solid line). 
(b) Corresponding measured 
$G_\parallel(r,t)$ and $G_\perp(r,t)$ (dashed lines),
compared with the prediction (solid lines) from 
ring kinetic theory in the incompressible limit.}
\end{figure}
\end{document}